\documentclass[11pt,letterpaper]{amsart}

\usepackage{amsmath, amssymb,latexsym, amsthm, amsfonts, amscd}
\usepackage{verbatim}
\usepackage{overcite}
\newtheorem{thm}{Theorem}

\pdfoutput=1

\usepackage{graphics}

\usepackage{citesort}
\usepackage{amssymb,amsmath}
\usepackage[mathscr]{eucal}

\normalsize

\title{Predicting Knot or Catenane Type of Site-Specific Recombination Products}

\author{Dorothy Buck and Erica Flapan}

\address{Department of Mathematics and Centre for Bioinformatics, Imperial College London, London England SW7 2AZ UK}
\email{d.buck@imperial.ac.uk}
\address{Department of Mathematics, Pomona College,
Claremont, CA 91711, USA}
\email{eflapan@pomona.edu}

\newcommand{\nibf}{\noindent \textbf}

\newcommand{\beq}{\begin{equation}}
\newcommand{\eeq}{\end{equation}}
\newcommand{\bea}{\begin{eqnarray}}
\newcommand{\eea}{\end{eqnarray}}

\begin{document}

\maketitle

%
%
%
%

\bigskip


\section{Summary}

Site-specific recombination on supercoiled circular DNA yields a
variety of knotted or catenated products. We develop a model of
this process, and give extensive experimental evidence that the
assumptions of our model are reasonable. We then characterize all
possible knot or catenane products that arise from the most common
substrates. We apply our model to tightly prescribe the knot or
catenane type of previously uncharacterized data.


\medskip

\textit{Keywords} Site-specific recombination, DNA knots, serine
recombinases, tyrosine recombinases, DNA topology {\footnote{\
Email address of the corresponding author: d.buck@imperial.ac.uk}}


%






\section{Introduction}

Since their discovery in the late 1960s, DNA knots and catenanes
(\textit{aka} links) have been implicated in a number of cellular
processes (see\ \cite{MB,Cozz} and references therein).
In particular, they arise during replication and recombination,
and as the products of enzyme actions, notably with
topoisomerases, recombinases and transposases\cite{Cozz,Path}.
The variety of DNA knots and catenanes observed has made
biologically separating and distinguishing these molecules a
critical issue.

Experimentally, DNA knots and catenanes can be resolved either via
electron microscopy or electrophoretic migration
\cite{KrasStas,Trig,ZechCris}.
Electron microscopy can definitively determine the precise knot or
catenane type; however this process can be both difficult --
particularly deciphering the sign of crossings -- and labourious.
Alternately, gel electrophoresis will stratify nicked DNA
knots/catenanes of a given molecular mass and charge. Typically,
the distance a given knot or catenane migrates through the gel is
proportional to the minimal crossing number (MCN, defined below),
with knots of greater MCN migrating more rapidly than those with
lesser MCN \cite{Stasiak,Sundin2,Levene} \footnote{\ However there
are gel conditions where, for example, the unknot will migrate
ahead of the trefoil.}
But there are 1,701,936 knots with MCN
$\leq$ 16, so a better stratification is needed to positively
identify a particular knot \cite{HosThis}. Recent work has shown
that 2-dimensional gel electrophoresis can separate some prime
knots with the same MCN \cite{Trig}. However, there is no clear
relationship that determines relative migration of knots with the
same MCN in the second dimension.

For DNA of a given length, (1-dimensional) gel electrophoresis can
separate some knots with the same MCN. For example, the 5 and
7-crossing torus knots migrate more slowly than the corresponding
5 and 7-crossing twist knots \cite{Stasiak,VologCri}. This has not
generalized, although recent experiments indicate that
knots/catenanes may migrate linearly with respect to the average
crossing number (ACN) of a particular conformation -- the
\textit{ideal configuration}\footnote{\ \textit{Ideal geometric
configurations} of knots or catenanes are the trajectories that
allow maximal radial expansion of a virtual tube of uniform
diameter centered around the axial trajectory of the knot
\cite{Gros,Katritch}.} of the knot or catenane
\cite{VologCri,Laurie,Katritch}.


For gel electrophoresis, one must also construct an appropriate
knot ladder as a control to determine the exact DNA knot or
catenane, since adjacent bands determine only relative MCN or ACN,
not precise values. While this can be done in some cases
(\textit{e.g.} T4 topoisomerase will produce a ladder of twist
knots \cite{WassCozz2}) generating such a ladder of known
knots/catenanes from DNA of the same length and sequence as the
unknown knots is highly nontrivial.



Thus topological techniques, such as those presented here, can aid
experimentalists in characterizing DNA knots and catenanes.

In this work, we focus on knots and catenanes that arise from
site-specific recombination. \textit{Site-specific recombinases}
mediate such a rearrangement of the genome (see \textit{e.g.}
\cite{MobDNA,Grindley6} for a more thorough introduction).
Loosely, the recombinases bind and synapse two small segments of
DNA, then cleave (by nucleophilc displacement of a DNA hydroxyl by
a protein side chain), exchange and reseal the backbones, before
releasing the DNA. The result of site-specific recombination can
be excision, integration, or inversion of DNA. This corresponds to
a wide variety of physiological processes, including integration
of viral DNA into the host genome, bacterial gene replication and
plasmid copy number control. If the substrate DNA contains
supercoils, or if synapsis introduces DNA crossings, these
crossings can become knot or catenane nodes in the product.

Topological techniques have already played a significant role in
characterizing knotted and catenated products of site-specific
recombination. For example, several approaches have been developed
to determine a particular DNA knot or catenane type, including
utilizing the node number for knots \cite{Cozz}, the Jones
polynomial for catenanes \cite{Boles},
Schubert's classification of 4-plats \cite{WhiteCozz} and the
HOMFLY polynomial \cite{WhiMilCozz}. Perhaps most famously, Ernst
and Sumners have developed the tangle model of recombination to
describe the action of particular site-specific recombinases in
terms of tangle sums \cite{ES1}.
The tangle model has since been
used to determine various features of protein-DNA interactions for
a number of specific proteins
\cite{us,us2,Cri,Dar,DarcyLevene2,me,ES2,SECS,Vaz,Vaz2}.


With the exception of \ \ \cite{SECS} discussed below, the
previous topological treatments began with the precise,
biologically determined knot or catenane types of (at least some
of) the products. This input was then harnessed in topological
arguments that probed various features of the pathway and/or
mechanism.
Here we consider the alternate paradigm: given a few assumptions
about the mechanism, we predict which knots/catenanes are putative
products.

More specifically, rather than focusing on a specific recombinase
as many earlier studies have done, we present a topological model
that predicts which knots/catenanes can occur as products of
site-specific recombination \textit{in general}. We do this by
describing the topology of how DNA knots and catenanes are formed
as a result of a single -- or multiple rounds of processive --
recombination event(s), given an unknot, unlink, or $(2,m)$-torus
knot or catenane substrate. Our model relies on three assumptions,
and we provide biological evidence for each. Given these
assumptions, this model predicts that products arising from
site-specific recombination must be members of a single family of
products (illustrated in Figure \ref{productfamily}).\footnote{\
Note \textit{all} figures represent (the axis of) duplex DNA.} In
\ \cite{BFmath} we provide the technical proofs for the model
developed here, whose nascent form we sketched in~\cite{OCAMI}.

This paper complements earlier work of Sumners \textit{et
al}~\cite{SECS}, which used the tangle model and several
biologically reasonable assumptions to solve tangle equations.
They then determined which 4-plat knots and catenanes arise as a
result of (possibly processive) site-specific recombination on the
unknot for the serine subfamily of recombinases. (See below for a
discussion of the 2 subfamilies.) For the particular case of the
recombinase Gin, they considered the knots $3_1,4_1,5_2$ or $6_1$
as substrates as well as unknotted substrates. The current work
goes further in several ways. In addition to an unknotted
substrate for a generic recombinase, we allow substrates that are
unlinks with one site on each component, as well as $(2,m)$-torus
knots and catenanes. Also, our assumptions are based exclusively
on the biology of the recombination process. In particular, we do
not assume the tangle model holds or that the products must be
4-plats. This is particularly important as recombination has been
seen to produce knots and catenanes which are connected sums, and
thus not 4-plats (see Table \ref{t:knownproducts}).

The paper is organized as follows. In Section 2, we state the
three assumptions of our model about the recombinase-DNA complex,
and give supporting experimental evidence for each. In Section 3,
we show that, given an unknot, unlink or $(2,m)$-torus knot or
catenane substrate, all possible knotted or catenated products
fall into a single characterized family. We also consider the
(common) case of substrates which are $(2,m)$-torus knots and
catenanes whose products have minimal crossing number $m+1$, and
show that the product knot or catenane type is even more tightly
prescribed. In Section 4, we first show that for substrates which
are unknotted, unlinked, or $(2,m)$-torus knots or catenanes all
known products fall within our family. (The technical proofs of
the results in Sections 3 and 4 can be found in \cite{BFmath}.) We
then apply our results to narrow the possible knot or catenane
type for previously uncharacterized experimental data.

\subsection{Background and Terminology}
\label{sec:bkgrd}

We define standard DNA as that which is covalently closed, duplex
and plectonemically supercoiled possibly with branch points.
Roughly speaking a circular DNA molecule is plectonemically
supercoiled if there is a second order helix formed by the DNA
axis itself (see \cite{MB} for a more complete description). It is
believed that supercoiled DNA is the typical form of DNA
\textit{in vivo}\cite{Boles}. Branched DNA structures within
supercoiled plasmids \textit{in vitro} have been visualized by
electron microscopy\cite{Adrian,Boles}. Additionally atomic force
microscopy \textit{in situ} illuminates branched plectonemic
superhelices at physiological conditions \cite{Lyub2}. \textit{In
vivo}, there is evidence from several more indirect experiments
that branched supercoiled DNA is ubiquitous, e.g.
\cite{MerJohnson}.


During site-specific recombination, a recombinase dimer first
binds to each of two specific DNA sites of approximately 20-30
basepairs. We refer to these sites as the \textit{crossover sites}
The two crossover sites are then brought
together within a \textit{recombinase complex},
$B$:
the smallest convex region containing the four bound recombinase
molecules and the two crossover sites. So $B$ is a {\it
topological ball} (\textit{i.e.}, it can be deformed to a round
ball). The crossover sites can be located either on the outside,
separated by the catalytic domains, (\textit{e.g.} with
$\gamma\delta$ and Tn3 resolvase), or inside the 4 recombinase
subunits
\cite{Grindley4,Nollmann,MerJohnson,Li,Ennifar,Guo,Rice,Biswas}.
We will use the term {\it recombinase-DNA complex} to refer to $B$
together with the substrate. If the recombinase complex meets the
substrate in precisely the two crossover sites then we say the
recombinase complex is a {\it productive synapse}.

\begin{figure}[htpb]
\includegraphics{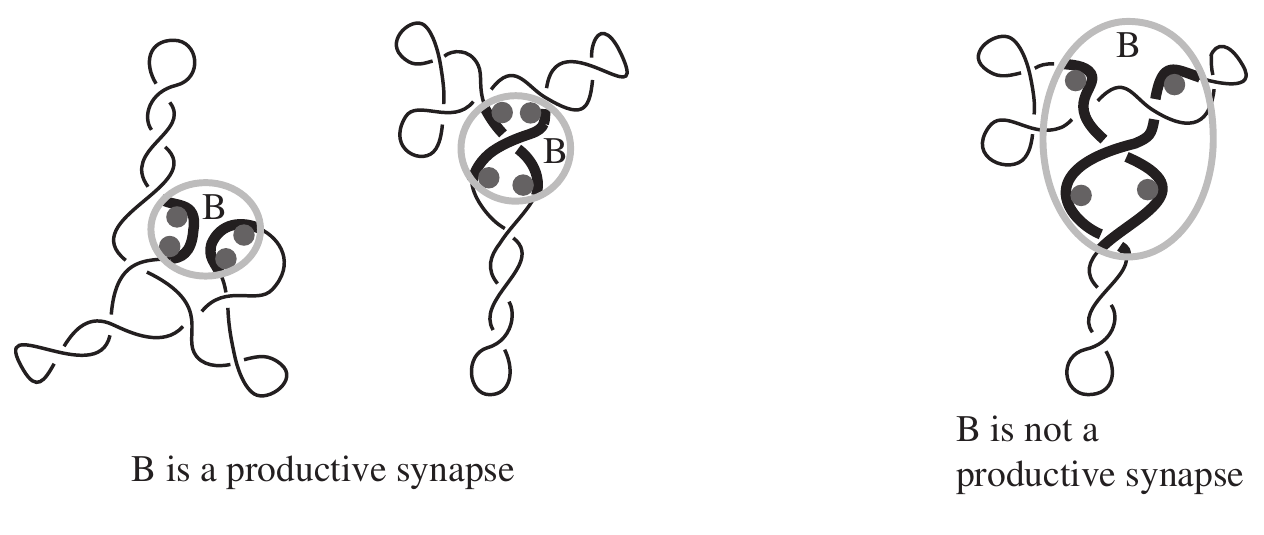}
\caption{Productive Synapse. We require that the recombinase
complex is a \textit{productive synapse}, that is the recombinase
complex meets the substrate in precisely the two crossover sites.
The two examples on the left have a productive synapse and the one
on the right does not. The crossover sites are highlighted in
black.} \label{productivesynapse}
\end{figure}

The existence of a productive synapse for recombinases is in
contrast with tranposases
whose enhancer sequences are intertwined with the active
transposition sites,\textit{e.g.}\cite{Path},
preventing the existence of a
productive synapse. Figure \ref{productivesynapse} demonstrates
two examples where the recombinase complex $B$ is a productive
synapse, and one where $B$ is not.

Site-specific recombinases fall into two families -- the serine
(also known as the resolvase) and tyrosine (also known as the
integrase) recombinases -- based on sequence homology and
catalytic residues \cite{Grindley6}.
The serine and tyrosine recombinases also differ in their
mechanism of cutting and rejoining DNA at the crossover sites.
Both families are large: a phylogenetic analysis has been
performed on 72 serine recombinases \cite{SmithThorpe} and a
recent iterative PSI-BLAST search documents approximately 1000
related sequences of putative tyrosine recombinases \cite{Alt}.

The large, diverse family of serine recombinases is comprised of
resolvases (such as Tn3 and $\gamma \delta$), invertases (such as
Gin, Hin, Pin, and Min), large serine recombinases (also called
large resolvases) and IS elements \cite{SmithThorpe}. These
recombinases may trap a fixed number of supercoils before
initiating recombination \cite{Grindley6}. For example, Tn3
resolvase requires three negative supercoils to be trapped by the
binding of (non-active) resolvase molecules. These trapped
supercoils (outside of the recombinase complex) together with the
recombinase complex itself are known as the \textit{synaptic
complex} \cite{WassDunCozz,Stark2}.
Likewise, the invertases also require a fixed number of supercoils
trapped outside the recombinase complex. Rather than using
additional recombinase molecules, they rely on accessory proteins
and enhancer sequences, which facilitate the organization of a
unique stereospecific synapse that promotes DNA cleavage. (In the
Hin and Gin systems, these bound supercoils, together with the
recombinase complex, are referred to as the \textit{invertasome}
\cite{Heich,MerJohnson}.) With serine recombinases, recombination
proceeds through a concerted 4-strand cleaving and rejoining
reaction\cite{Grindley6}.
Serine recombinases can perform multiple rounds of strand exchange
before releasing the DNA, in a process known as \textit{processive
recombination}.

In contrast, tyrosine recombinases first cleave, exchange and
reseal two sugar-phosphate backbones. The DNA-protein complex then
proceeds through an intermediary structure (a Holliday junction)
before repeating the process with the other two DNA backbones
\cite{GJ,Grindley6}.
Most tyrosine recombinases, including Flp, $\lambda$ Int and Cre,
tolerate varying number of supercoils outside of the recombinase
complex. However, there are exceptions, most notably XerCD, which
trap a fixed number of supercoils using accessory proteins before
initiating cleavage \cite{Colloms}. Like serine recombinases,
tyrosine recombinases can also employ accessory proteins to help
assemble the synaptic complex, and to drive the overall reactions
(\textit{e.g.} $\lambda$ Int and XerCD) \cite{Colloms,Biswas}.

\smallskip

Finally, we shall use the following knot theoretic terms. The {\it
components} of a catenane are the separate rings of the catenane.
A knot is considered to be a catenane with only one component. A
$(2,m)$-torus knot or catenane is one which can be drawn so that
all of its crossings occur as a row of $m$ twists, as illustrated
in Figure \ref{T(2,m)Paper}. We will denote a knot or catenane of
this form by $T(2,m)$. Any such knot or catenane is the boundary
of a twisted annulus. If $m$ is odd then $T(2,m)$ is a knot and if
$m$ is even then $T(2,m)$ is a catenane. Finally, given two knots
or catenanes $K$ and $J$, their \textit{connected sum}, written $K
\# J$, is obtained by removing a trivial arc from each and gluing
the resulting two endpoints of $K$ to the two endpoints of $J$
without introducing any additional knotting. Figure
\ref{productfamily}, subfamilies 4 and 5 give examples of
connected sums.

\begin{figure}[htpb]
\includegraphics{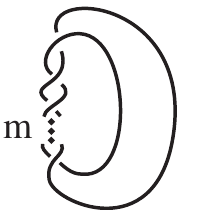}
\caption{Torus Knots and Catenanes. A $(2,m)$-torus knot or
catenane has this form. The model considers substrates that are
unknots, unlinks or $(2,m)$-torus knot or catenanes.}
\label{T(2,m)Paper}
\end{figure}

\section{The assumptions of our model}

We begin with a fixed recombinase and an unknot, unlink, or $T(2,m)$ substrate. If
the substrate is an unlink then we assume that one site is on each
component, as otherwise this case reverts back to a single
unknotted substrate. We make three biological assumptions about
the recombinase-DNA complex, stated in both biological and
mathematical terms, and provide experimental evidence for each.

Let $J$ denote the substrate(s) after synapsis, and recall that
$B$ denotes the recombinase complex.

\medskip


\nibf{Assumption 1:} The recombinase complex is
a productive synapse, and there is a projection of the crossover
sites which has at most one crossing between the sites and no
crossings within a single site.

\medskip

\nibf{Evidence for Assumption 1:}


We present a variety of experimental data that suggests that the
recombinase complex is sufficiently dense both to form a proper
productive synapse and to preclude extraneous crossings.

Most convincingly, recent crystal structures of several
recombinase complexes support both of these assumptions
\cite{Biswas,Ennifar,Guo,Rice2,Rice}. Additionally, structures of
a single site synapsed with either a dimer or monomer indicate
that there are no crossings within an individual
site\cite{Blake,MacDonald,Yang}. Furthermore, structures of two
intermediate complexes -- a synaptic $\gamma \delta$ resolvase
tetramer covalently linked to two \textit{cleaved} DNAs, and the
Flp recombinase-Holliday junction complex have at most one
crossing between sites and none within a site \cite{Li,Rice}. Thus
the large-scale conformational changes necessary to unwrap
crossings during the reaction imply it is unlikely that the
crossover sites contain additional crossings at synapsis or that a
productive synapse does not exist.

Also, there are significant DNA-protein and protein-protein
interactions that appear to prevent additional crossings and
extraneous strands from piercing the recombinase complex. With
tyrosine recombinases, each domain flanking the crossover site DNA
inserts a helix into a major groove, and the highly conserved
C-terminal domain interacts with consecutive minor and major
grooves on the opposite face of DNA \cite{Grindley6}. With serine
recombinases, DNA binding involves the conserved H-T-H domain, and
a DNA binding domain on the C or N terminus of the protein
\cite{Li,Yang,SmithThorpe}. For both families, there are also
significant protein-protein interfaces with the other proteins in
an assembled tetrameric complex. Additionally, DNA itself has a
geometric diameter of 2nm, and, depending on the ionic conditions,
a much greater electrostatic diameter (\textit{e.g.} $\simeq$ 5nm
at physiological conditions)
\cite{VologCozz}.


Additional biochemical experiments support both the existence of a
productive synapse and a bound on the crossings between or within
the sites. Atomic force microscopy of both the Cre and Flp
productive synapses concur with the conclusions drawn from the
crystal structures of the resolvases and integrases
\cite{DarcyLevene2}. Also the architecture of the $\gamma \delta$
resolvase recombinase complex has been determined to be a
productive synapse with a single crossing, in experiments using
constrained DNA \cite{Grindley4}. Furthermore, solution structures
from neutron and Xray scattering data of hyperactive Tn3 resolvase
mutants show that a productive synapse exists, and that there is a
projection of the sites with at most one crossing between sites
and no crossings within a single site \cite{Nollmann}.
Additionally, recent cyclization experiments indicate that dimers
of Flp and Cre each bend the DNA sites upon binding, but not
enough to introduce a crossing within a single site \cite{Du}. The
steric and electrostatic constraints mentioned above imposed by
the short length of the sites also putatively limit crossings
between and/or within the sites.

Finally, we note that for recombinases that utilize accessory
proteins, we recall that an accessory site or enhancer sequence is
neither a crossover site nor a part of a crossover site. Thus in
order for our assumption to hold, if a recombinase requires an
enhancer sequence, then it must be sequestered from the crossover
sites. In particular, we claim that when the enhancer or an
accessory site loops around to form a specific recombinase complex
all crossings are trapped outside of the complex, even though the
recombinases might interact directly with the enhancer sequence.
Supporting evidence is twofold. Firstly, the recent $\lambda$
Int-DNA complex crystal structure includes the accessory sites,
and it is clear that a productive synapse exists and has the
required limited number of crossings \cite{Biswas}. Additionally,
support for the invertase family comes both from detailed
biochemical experiments of the Hin system. The standard molecular
model of the Hin invertasome, based on the cross-linked structure
of the Hin-DNA co-complex, has two Fis dimers bound to the
enhancer sequence and two Hin dimers bound at the recombination
sites \cite{MerJohnson}.
According to this model, the enhancer sequence is sequestered from
the crossover sites, and the crossover sites are not interwound.





All of the above evidence indicates that it is biologically reasonable to
assume that a given recombinase-DNA complex satisfies
Assumption~1.

\medskip

\noindent{\bf Assumption 2:} The productive synapse does not
pierce through a supercoil or a branch point in a nontrivial way.
Also, no persistent knots are trapped in the branches of the DNA
on the outside of the productive synapse.

%
%
%
%
%
%
%
%

\begin{figure}[htpb]
\includegraphics{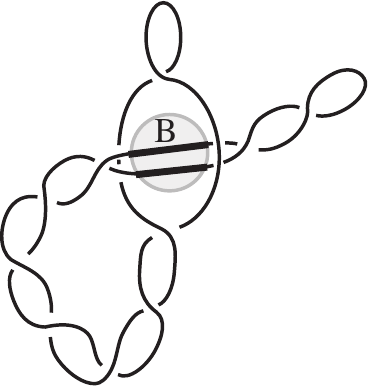}
\includegraphics{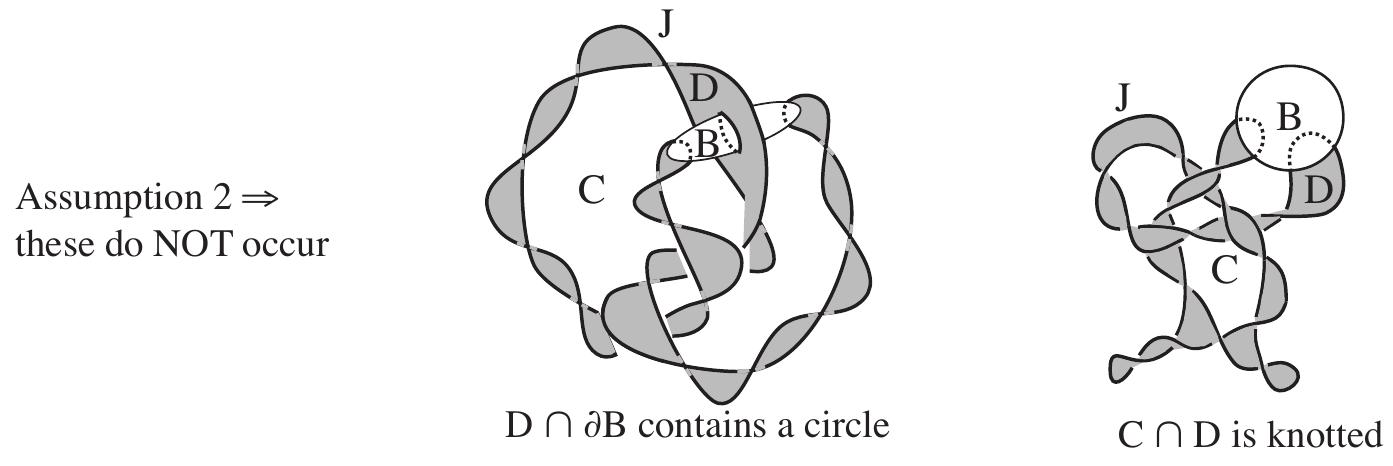}
\includegraphics{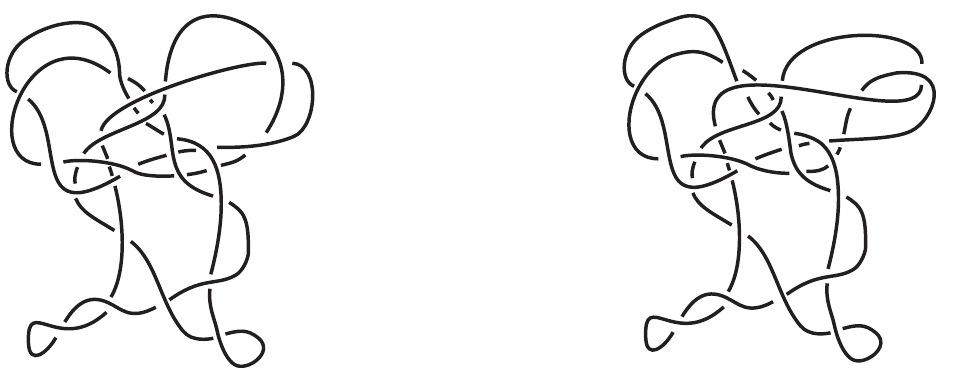}
\caption{Scenarios for Assumption 2. \textit{Above:} the
productive synapse $B$ trivially pierces through a supercoil. This
is allowed. \textit{Middle Left:} $B$ pierces through a supercoil
in a nontrivial way. \textit{Middle Right:} A knot is trapped in
the branches on the outside of $B$. These are forbidden.
\textit{Below:} If knots were trapped within the branches of the
substrate(s) on the outside of the productive synapse, then
recombination would at least occasionally produce {\it satellite}
knots and catenanes such as these. These are forbidden.}
\label{Assumption2}
\end{figure}

\medskip

\noindent{\bf{Evidence for Assumption 2:}}

A variety of microscopy studies support Assumption 2. Atomic force
microscopy revealed that at physiological conditions, supercoiled
DNA adopts a compact plectonemic configuration with close
juxtaposition of DNA segments in the loops, which makes it
unlikely that a supercoiled domain could be penetrated
\cite{Lyub2,Shlyak}.
Under conditions that minimize intersegmental repulsion, electron
and scanning force microscopy studies both demonstrate that
opposing segments of interwound supercoiled DNA are frequently
close together \cite{Adrian,Boles,Cherny}.
Probabilistically then, it is thus unlikely that either a
supercoiled domain or $B$ could pierce through a supercoil.

Also, experimental work coupled with MMC simulations indicate that
on average supercoiled DNA helices are separated by 10nm -- only 5
times the width of the DNA diameter itself \cite{Boles,VologCozz}.
Since, as discussed above, the \textit{electrostatic} diameter of
DNA in physiological conditions is closer to 5nm, it seems quite
unlikely that the productive synapse could pass through a
supercoil (as in Figure \ref{Assumption2})
\cite{Ryben1,Ryben2,Yarmola}. Brownian dynamics simulations of
site juxtaposition support these findings \cite{Huang3}.

Additionally, the probability of one duplex (linear) invading a
supercoiled domain has been shown to be quite low, by both
experiments and MMC simulations \cite{Ryben1}. This frequency may
be even lower if a supercoiled domain, rather than linear duplex,
invades another supercoiled domain. Therefore trapping a
persistent knot within the branches of the DNA during synapsis is
also unlikely.

Also the steric and electrostatic constraints arising from
protein-DNA interactions discussed in the evidence for Assumption
1 would appear to preclude piercing of the productive synapse by
non-site DNA.

Finally, if persistent knots could be trapped within the branches
of the substrate(s) on the outside of the productive synapse then
we would expect to see (at least occasionally) ``doubly knotted"
products like those illustrated in Figure \ref{Assumption2} (such
knots and catenanes are known as {\it satellites}). However, no
products like these have thus far been observed (see Table
\ref{t:knownproducts}). This indicates that knotting of the
branches is unlikely to occur.

All of the above evidence indicates that it is biologically
reasonable to assume that a given recombinase-DNA complex
satisfies Assumption~2.

\medskip

Next, we state Assumption~3, which addresses the mechanism of recombination for serine
and tyrosine separately.

\medskip

\noindent {\bf  Assumption 3 for Serine Recombinases:}
Serine recombinase performs recombination via the ``subunit
exchange mechanism." This mechanism involves making two
simultaneous (double-stranded) breaks in the sites, rotating
opposites sites together by $180^{\circ}$ within the productive
synapse and resealing opposite partners. In processive
recombination, each recombination event is identical.
\medskip

\nibf{Assumption 3 for Tyrosine Recombinases:} After recombination
mediated by a tyrosine recombinase, there is a projection of the
crossover sites which has at most one crossing.

\medskip

\noindent{\textbf{Evidence for Assumption 3:}}

\noindent{\textit{Serine Recombinases:} A large number of
\textit{in vitro} topology studies performed on DNA invertases and
resolvases have provided solid support for the ``subunit exchange
mechanism,'' where one set of recombinase subunits, each
covalently associated with the 5' ends of the cleaved
recombination sites, switches places, resulting in a 180$^\circ$
rotation of DNA strands. (See \ \cite{Grindley6} and references
therein.)
This is supported by a recent crystal structure of a synaptic
tetramer of $\gamma \delta$ resolvase covalently catenated to two
cleaved DNA molecules, indicating a subunit rotation of $180
^\circ$ \cite{Li}.

Additional experiments involving Tn3, Hin and Gin lends credence
to the idea that each round of processive recombination acts
identically \cite{Cozz,Kanaar,Heich}. For example, Heichman
\textit{et al} demonstrate that
there are multiple rounds of exclusively clockwise subunit
rotation of one set of Hin subunits after DNA cleavage
\cite{Heich}.

\smallskip

\noindent{\textit{Tyrosine Recombinases:} While there are no known
post-recombinant crystal structures, there are synaptic
intermediary crystal co-complexes for Flp \cite{Rice}, Cre
\cite{Guo}, and $\lambda$ Int \cite{Biswas} (this also including
accessory sites in addition to the typical $\lambda$ Int crossover
sites). These structures indicate that at the earlier stages of
recombination -- namely after the first cleavage, exchange or
within a Holliday junction intermediate -- there exists a
projection with at most one crossing. They also highlight
particular features of the productive synapse that may impede the
large-scale conformational changes needed to introduce crossings.

As mentioned above, the protein-DNA interface is a large
hydrogen-bonded network. Flp, Cre and $\lambda$ Int all form a
C-shaped clamp around the DNA substrate, and the C-terminal
domains interact with consecutive minor and major grooves on the
opposite face of the DNA \cite{Grindley6}. Additionally, there are
significant protein-protein interactions, \textit{e.g.} the
catalytic domains interact by swapping part of the C-terminus with
a neighbouring protomer.

Also, the post-recombinant complex is formed from the Holliday
junction intermediate by, first an isomerization of the
intermediary complex so that the inactive monomers become active
and \textit{vice versa}, and then a repeated strand cleavage where
the new 5' ends migrate over and attack their partners' 3'
phosphotyrosine linkages. This second round of strand transfer
completes the reaction. Particularly given the two-fold symmetry
of the reaction, it thus seems unlikely that in the final stage of
recombination there is enough motion of the DNA arms to generate
multiple additional crossings between sites or a crossing within a
single site.

\textit{In vitro} studies also suggest that tyrosine recombinases
that mobilize the gene cassettes of integrons may preferentially
bind DNA hairpins, which would constrain the number of crossings
\cite{Johansson}.
Finally, given the steric
and electrostatic constraints of short DNA arms discussed for
Assumption 1, it is probable that there exists a projection of the
sites containing at most one crossing between sites and no
crossings of a single site within the post-recombinant complex.

\smallskip

All of the above evidence indicates that it is biologically
reasonable to assume that a given recombinase-DNA complex
satisfies Assumption~3.

\section{Results}

\subsection{All products of unknots,unlinks or $(2,m)$ torus links substrates fall within a single family.}

In this section, we suppose that the substrate is an unknot, an
unlink, or $T(2,m)$ and that all three of our assumptions hold for
a particular recombinase-DNA complex. Below we state Theorems
\ref{T:tyrosine} and \ref{T:serine}, whose technical proofs can be
found in \cite{BFmath}. These theorems demonstrate that all
knotted and catenated products brought about by that recombinase
are in the family of knots and catenanes illustrated in Figure
\ref{productfamily}.

\begin{figure}[h]
\includegraphics{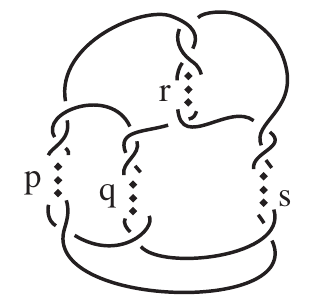}
\includegraphics{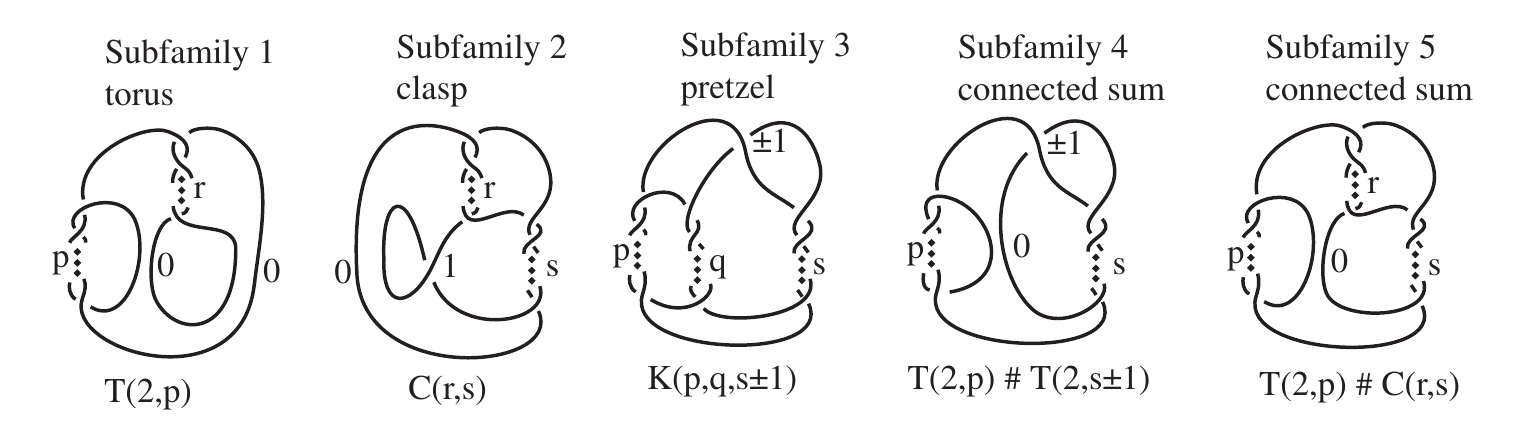}
\caption{Product Family. Given an unknotted, unlinked or torus
knot or catenane substrate, then all products are of this type.
The five subfamilies are indicated below.} \label{productfamily}
\end{figure}

Observe that $p$, $q$, $r$, and $s$ can be positive, negative, or
zero. Furthermore, by letting $p$, $q$, $r$, and/or $s$ be 0 or
$1$ in Figure \ref{productfamily} as appropriate, we obtain the
five subfamilies illustrated in Figure \ref{productfamily}.
Observe that if $q=0$, $r=1$, and $s=-1$, then we have a $T(2,p)$
together with an uncatenated trivial component. This possibility
occurs as a member of Subfamily 4. Thus the knots and catenanes in
these subfamilies are all possible products of recombination as
specified in Theorems 1 and 2. We use the notation $C(r,s)$ for a
knot or catenane consisting of one row of $r$ crossings and a
non-adjacent row of $s$ crossings (illustrated in Subfamily 2).
Note that if $r$ or $s$ equals 2, then $C(r,s)$ is in the well
known family of {\it twist} knots and catenanes. We us the
notation $K(p,q,r)$ for a pretzel knot or catenane with three
non-adjacent rows containing $p$ crossings, $q$ crossings, and $r$
crossings (illustrated in Subfamily 3, where $r=s\pm1$). Note, by
{\it non-adjacent} rows of $r$ and $s$ crossings we mean that the
two rows cannot be considered as a single row of $r+s$ crossings.

\begin{thm}\label{T:tyrosine}
Suppose that Assumptions 1, 2, and 3 hold for a particular
tyrosine recombinase-DNA complex. If the substrate is an unknot
then the only non-trivial products are $T(2,n)$ and $C(2,n)$. If
the substrate is an unlink, then the only non-trivial product is a
Hopf link. If the substrate is $T(2,m)$ then all of the
 non-trivial products are contained in the family illustrated in Figure \ref{productfamily}.
\end{thm}

\medskip

\begin{thm}\label{T:serine}
Suppose that Assumptions 1, 2, and 3 hold for a particular serine
recombinase-DNA complex. If the substrate is an unknot then the
only non-trivial products are $T(2,n)$ and $C(p,q)$. If the
substrate is an unlink, then the only non-trivial product is
$T(2,n)$. If the substrate is $T(2,m)$ then all non-trivial
products are in the general family illustrated in Figure
\ref{productfamily}.
\end{thm}
\medskip

Table 1 summarizes the non-trivial products predicted by Theorems
1 and 2. Recall that $C(r,s)$ is a twist knot or catenane if $r$
or $s$ equals 2.

{\begin{table}
    \begin{tabular}{||l|c|c||} \hline \hline

Recombinase Type & Substrate Topology &   Non-trivial Products \\
         \hline \hline
         Tyrosine & unknot & $T(2,n)$, $C(2,n)$ \\
        \hline
    & unlink & Hopf link$=T(2,2)$\\
        \hline
    & $T(2,m)$ & Family of Figure \ref{productfamily} \\
        \hline
Serine & unknot & $T(2,n)$, $C(p,q)$\\
        \hline
    & unlink  & $T(2,n)$\\
        \hline
    &  $T(2,m)$ & Family of Figure \ref{productfamily} \\
  \hline

\end{tabular}
\label{t:predictions}

\medskip

\caption{Non-trivial products
predicted by our model.}

\end{table}

\subsection{Products are more tightly restricted when recombination
adds 1 crossing}

Knots and catenanes have been tabulated according to the fewest
number of crossings with which they can be drawn (see the tables \
\cite{Rolf} and \ \cite{HosThis}). This number of crossings is
called the {\it minimal crossing number} of the knot or catenane,
and is denoted by MCN. For example a $T(2,2)$ (also known as a
Hopf link) has MCN$=2$ and $T(2,3)$ (also known as a trefoil knot)
has MCN$=3$. In fact, $\mathrm{MCN}(T(2,m)=m$ for any positive
integer $m$. Gel electrophoresis can be used to determine the MCN
of a product (see \textit{e.g.} \cite{Levene}).

It is often the case that recombination adds a single crossing to
the MCN of a knotted or catenated substrate, \textit{e.g.}
\cite{Bath}. If the substrate is $T(2,m)$ and the product has
$\mathrm{MCN}=m+1$, then we can further refine the results of
Theorems 1 and 2 to determine more specific possibilities for the
products, the technical details of which can be found in
\cite{BFmath}. The conclusion of the theorem is illustrated in
Figure \ref{mcntheorem}.
\medskip

\begin{thm}\label{T:MCN}
Suppose that Assumptions 1, 2, and 3 hold for a particular
recombinase-DNA complex with substrate $T(2,m)$, with $m > 0$. Let
$L$ be the product of a single recombination event, and suppose
that $\mathrm{MCN}(L)=m+1$. Then $L$ is either $T(2,m+1)$,
$C(m-1,-2)$, or $K(p,q,1)$ with $p,q>0$ and $p+q=m$.

Furthermore, $K(p,q,1)$ is a knot if and only if at least one of
$p$ and $q$ is odd.

\end{thm}
  \begin{figure}[htpb]
\includegraphics{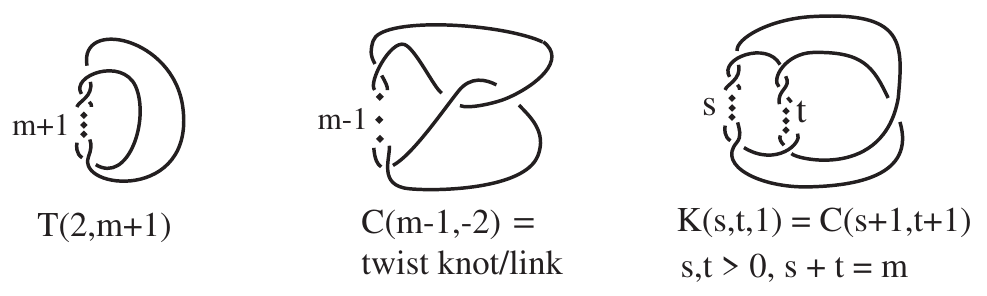}
\caption{Restricted Products. These are the only possible
products, if the substrate is $T(2,m)$ and the product has
MCN$=m+1$.} \label{mcntheorem}
\end{figure}


\section{Applications}

\subsection{All Characterized Recombinant Products are in the
Predicted Family}

Table 3 summarizes the known products of recombinases starting
with substrates which are unknots, unlinks, or $T(2,m)$. As shown
in Table 3, all products listed have a projection in the form of Figure
\ref{productfamily}. This provides further confirmation of the
validity of our model.

Note that this table does not describe every product of
site-specific recombination -- e.g. Tn3 acting on the twist knot
$4_1$ yields the product $5^2_1$ \ \cite{WassDunCozz}, and mutant
Hin acting on a heterogenous population of $3_1$ and twist knots
$5_2$ yields (double and, in multiple rounds of recombination,
triple) connected sums: $3_1\# 3_1, 3_1\# 5_2, 5_2\# 5_2$ and
$3_1\#3_1\#5_2$ \ \cite{MerJohnson} -- but we do not consider
these (twist knot or two separate knots) as substrates in our
model.

\begin{footnotesize}
\begin{table}
    \begin{tabular}{||c|l|c|c|l||} \hline \hline

Recombinase & Substrate & Products & Subfamily & Reference\\
         \hline \hline

Cre & $0_1$ (Inverted) & $0_1, 3_1,5_1,7_1,9_1, 11_1$ &  1 & \cite{Hoess}\\
        \hline
    & $-2^2_1$ (Direct) & $0_1,3_1$ &  1 & \cite{Cri}\\
        \hline
    & $0_1$ (Direct) & $0^2_1,2^2_1,4^2_1,6^2_1,8^2_1,10^2_1,12^2_1$ &  1 & \cite{Hoess}\\
        \hline
Flp & $0_1$ (Inverted) & $0_1, 3_1,5_1,7_1,9_1, 11_1$ & 1 &
\cite{Cri,Cox}\\ \hline
    & $0_1$ (Direct) & $0^2_1,2^2_1,4^2_1,6^2_1,8^2_1,10^2_1,12^2_1$ &  1 & \cite{Cri,Cox}\\
        \hline
$\lambda$ Int & $0_1$ (PB inverted)  &  $0_1, 3_1,5_1,7_1,9_1, 11_1,13_1,15_1,17_1,19_1$ &  1 & \cite{Speng}\\
        \hline
              & $0_1$ (PB direct)  & $4^2_1,6^2_1$ &  1 & \cite{Speng}\\         \hline
              & $-2^2_1$ (PB direct)  & $5_2,7_2,9_2,11_2$ &  $2=C(2,s)$ & \cite{Cri}\\
        \hline
              & $0_1$ (LR inverted) & $3_1, 5_1, 7_1$ &  1 & \cite{Cri}\\
        \hline
              & $0_1$ (LR direct)  & $4^2_1$ &  1 & \cite{Cri}\\
        \hline
              & $-2^2_1$ (LR direct) & $0_1,3_1$, $5_2,7_2,9_2,11_2,13_2$ &  1, 2=$C(2,s)$ & \cite{Cri}\\
        \hline

Xer & $0_1$ (Direct) & $4^2_1$ &  1 & \cite{Colloms}\\
        \hline

        \hline

        \hline

Gin (inverted)  & $0_1$ & $0_1$ &  1 & \cite{Kanaar}\\
        \hline
& $0_1$ & $3_1$ & 1 & \cite{Kanaar} \\
        \hline
& $3_1$ & $4_1$ & 1 & \cite{Kanaar} \\
        \hline
& $3_1$ & $3_1\# 3_1$ & 4 & \cite{Kanaar}\\
        \hline
& $4_1$ & $5_2$ & 1 & \cite{Kanaar} \\
        \hline
Gin (direct) & $0_1$ & $3_1$ & 1 & \cite{Kanaar} \\
        \hline
& $3_1$ & $5_2$ & $2$ & \cite{Kanaar} \\
        \hline
& $3_1$ & $3_1\# 3_1$ & 4 & \cite{Kanaar}\\
        \hline
& $5_2$ & $7_2$ & $2$ & \cite{Kanaar} \\
        \hline

Gin (mutant, inv'd) & $0_1$ & $3_1,4_1,5_1,5_2,7_1,7_2,9_1,9_2$ &
1,2 &
\cite{Cris2}\\
        \hline

Hin (inverted) & $0_1$  & $0_1$ & 1 & \cite{MerJohnson} \\
        \hline
    & $0_1$ & $3_1$ & 1 & \cite{MerJohnson}\\
        \hline
    & $3_1$  & $4_1$ &  $2$ & \cite{MerJohnson} \\
        \hline
    & $4_1$  & $5_2$ &  $2$ & \cite{MerJohnson} \\
        \hline

Hin (mutant) & $0_1$ & $3_1,5_2, 3_1 \# 3_1$ & 1,2,4 & \cite{Heich}\\
        \hline
Tn3, $\gamma \delta$ (direct) & $0_1$ & $2^2_1$&  1 & \cite{WassDunCozz}\\
        \hline
    & $2^2_1$ & $4_1$ &  $2$  & \cite{WassDunCozz}\\
        \hline
    & $4_1$  & $5^2_1$ &  $2$ & \cite{WassDunCozz}\\
        \hline
    & $5^2_1$  & $4_1 \# 2_1$ &  4 & \cite{KrasStas}\\
        \hline
    & $5^2_1$  & $6_2$ &  $2$ & \cite{WassDunCozz}\\
        \hline
\end{tabular}

\medskip

\caption{As predicted by our model, all characterized products of
site-specific recombination on supercoiled unknotted, unlinked or
$(2,n)$-torus knot or catenane substrates fall within our single
family (see Figure~\ref{productfamily}).}

\label{t:knownproducts}

\end{table}
\end{footnotesize}

\subsection{Applications to Uncharacterized Recombinant Products}

We now turn our attention to several recombination systems whose
products are unclassified beyond minimal crossing number. We use
our model, together with results about minimal crossing number, to
prove that the product knot or catenane type is tightly
prescribed, and apply this new result to the previously
uncharacterized experimental data.

For each, we discuss how our model can help to restrict the knot
types of these products.

\medskip

\nibf{Xer:} Using a plasmid with both $\lambda$ Int and Xer sites,
Bath \textit{et al} generated the catenanes $6^2_1$ and $8^2_1$ as
products of $\lambda$ recombination \cite{Bath}. These were then
used as the substrates for Xer recombination, yielding a knot with
MCN=7 and a knot with MCN=9, respectively. These products have not
been characterized beyond their minimal crossing number. There are
seven knots with MCN=7 and 49 knots with MCN=9.

Theorem \ref{T:MCN} significantly reduces the number of
possibilities for each of these products. In particular, it
follows from Theorem \ref{T:MCN} that the 7-crossing products of
Xer must be $7_1=T(2,7)$, $7_2=C(5,-2)$ or $7_4=K(3,3,1)$; and the
9-crossing products of Xer must be $9_1=T(2,9)$, $9_{2}=C(7,-2)$,
or $9_5=K(5,3,1)$. Observe that all of these knots are 4-plats.
This demonstrates how our model complements earlier work of
\cite{Dar}, which assumes all products must be 4-plats and hence
only considers 7-crossing products (since only half of the
9-crossing knots are 4-plats). In \ \cite{new} we use our model
together with tangle calculus to completely classify all tangle
solutions to these $\lambda$ Int-Xer equations.


\medskip

\nibf{Cre111:} Abremski and collaborators created the mutant
Cre111, which yields products topologically distinct from those of
wild-type Cre \cite{Abremski}. When Cre111 recombines a
supercoiled substrate, the knotted and catenated products are, in
their conditions, significantly more complex than those produced
by wild-type Cre.

These knots and catenanes have thus far been uncharacterized.
However, our Theorem \ref{T:tyrosine} predicts that these knots
and catenanes must be of the form $T(2,n)$ or $C(2,n)$. Thus, by
running these products adjacent to a ladder of torus knots of the
same length, one could determine the exact knot or
catenane type.

\smallskip

%
%

\nibf{Tn3:} Benjamin \textit{et al} constructed a plasmid
substrate for Tn3 resolvase with four directly repeated crossover
sites \cite{Benjamin}. After the first round of recombination,
electron microscopy reveals the Hopf link $T(2,2)$ as the primary
product. After recombination, products were determined (via high
resolution gel electrophoresis (of 7-8 days) followed by electron
microscopy, as $T(2,2)$) \ $T(2,2) \# T(2,2)$ and two distinct
4-component catenates. Our model predicts that, with this 4-sited
substrate, recombination must proceed from the unknot to the Hopf
link $T(2,2)$. It then utilizes this $T(2,2)$ catenane as a
substrate to yield the product the connected sum of two Hopf links
$T(2,2) \# T(2,2)$ (see Table \ref{t:predictions}). This connect
sum is then the substrate for the products of 4-component
catenanes, but is not one of the substrates that we consider. This would be akin
to Tn3 performing multiple rounds of \textit{distributive}
recombination on a substrate with only 2 crossover sites. The
current work thus supports Benjamin \textit{et al}'s hypothesis of
neighbouring-site recombination.


\section{Concluding Remarks}

In this paper we have developed a model of how DNA knots and
catenanes are produced as a result of a recombinase acting on an
unknot, unlink, or $(2,n)$-torus knot or catenane substrate. Our
model is based on three explicitly stated biological assumptions
about site-specific recombination, and we have provided biological
evidence for each. It follows from our model that all knotted or
catenated products of such enzyme actions will be in the family of
Figure \ref{productfamily}, as described in Theorems
\ref{T:tyrosine} and \ref{T:serine}.

As mentioned above, the minimal crossing number (MCN) of a DNA
knot or catenane can be determined experimentally \cite{Levene}.
For small values of the MCN there are not many knots or catenanes
with a given value. However, the number of knots and catenanes
with MCN $=n$ grows exponentially as a function of $n$ \cite{ES3},
and there are 1,701,936 knots with MCN $\leq$ 16 \cite{HosThis}.
So knowing the MCN is not sufficient to determine the knot or
catenane.

However the total number of knots and catenanes in the family of
Figure \ref{productfamily} grows linearly with $n^3$
\cite{BFmath}. So the proportion of all knots and catenanes which
are contained in our family decreases exponentially as $n$
increases. Thus, knowing the MCN of a product and knowing that the
product is in one of our families allows us to significantly
narrow the possibilities for its knot or catenane type. The model
described herein thus provides an important step in characterizing
DNA knots and catenanes which arise as products of site-specific
recombination.

\section{Acknowledgements}

The authors wish to thank J. Arsuaga, I. Grainge, R. Harshey, M.
Jayaram, A. Stasiak, D.W. Sumners, A. Vologodskii and S.
Whittington for helpful conversations. DB was partially supported
by NSF Grant \# DMS-0102057. EF was partially supported by an
Association for Women in Mathematics Michler Collaborative
Research Grant.

\bibliographystyle{unsrt}
\bibliography{MasterBibFile}

\end{document}